\documentclass[%
 reprint,
 amsmath,amssymb,
 aps,
]{revtex4-1}

\usepackage{graphicx}
\usepackage{dcolumn}
\usepackage{bm}
\usepackage{amsfonts}
\usepackage{amssymb}
\usepackage[latin1]{inputenc}
\usepackage{textcomp}
\usepackage{epstopdf}

\begin{document}


\title{Self-trapped quantum walks}

\author{A. R. C. Buarque and W.S. Dias}
\affiliation{
Instituto de F\'isica, Universidade Federal de Alagoas, 57072-900 Macei\' o, Alagoas, Brazil
}%

\begin{abstract}
We study the existence and charaterization of self-trapping phenomena in discrete-time quantum walks. By considering a Kerr-like nonlinearity, we associate an acquisition of the intensity-dependent phase to the walker while it propagates on the lattice. Adjusting the nonlinear parameter ($\chi$) and the quantum gates ($\theta$), we will show the existence of different quantum walking regimes, including those with travelling soliton-like structures or localized by self-trapping. This latter scenario is absent for quantum gates close enough to Pauli-X. It appears for intermediate configurations and becomes predominant as quantum gates get closer to Pauli-Z. By using $\chi$ versus $\theta$ diagrams, we will show that the threshold between quantum walks with delocalized or localized regimes exhibit an unusual aspect, in which an increment on the nonlinear strength can induce the system from localized to a delocalized regime.
\end{abstract}
\maketitle

\section{introduction}
\label{introduction}

Quantum-mechanical systems in which the effective evolution is governed by a nonlinear equation are present in many branches of science, such as optics~\cite{nonlinear_optics,Review_mod_phy_nonlinear_optics,physreports_soliton_optics}, biology~\cite{Davydov_book,Phys_repo_soliton_davydov}, Bose-Einstein condensates~\cite{rev_mod_phys_Pitaevskii,rev_mod_phys_soliton,rev_mod_phys_BEC} and solid state physics~\cite{Holstein_1,Holstein_2,ssh_prl,rev_mod_phys_ssh}. In optical media, for example, nonlinearity arises from the field-induced changes in the refractive index of the propagation medium~\cite{nonlinear_optics,Review_mod_phy_nonlinear_optics,physreports_soliton_optics}, while for Bose-Einstein condensates the nonlinearity is related to interatomic interactions~\cite{rev_mod_phys_Pitaevskii,rev_mod_phys_soliton,rev_mod_phys_BEC}. Nonlinearity also appears as a result of lattice vibrations in the dynamic description of elementary excitations~\cite{Davydov_book,Phys_repo_soliton_davydov,Holstein_1,Holstein_2,rev_mod_phys_ssh,ssh_prl}.

Among the most interesting subjects related to nonlinearity are the self-trapping states. When associated to delocalized modes, initial  excitations display as signature a propagation without spreading (shape preserving), due to balancing between nonlinearity and linear correlation (dispersion, diffraction, diffusion) effects~\cite{nonlinear_optics,physreports_soliton_optics,Davydov_book,Phys_repo_soliton_davydov, rev_mod_phys_soliton,rev_mod_phys_BEC,Holstein_1,Holstein_2,ssh_prl,rev_mod_phys_ssh,Rev_mod_soliton_op_commu,
PRA_op_comm,PRA_op_comm_2,scien_op_comm,Buarque_nl,Hegger_rev,ssh_nature,SSH_polaron}. However, the absence of propagation are also a remarkable effect of self-trapping states. In this case, an initial  excitation is induced to trapping, with a significant time-averaged probability of finding it in a finite region of system when the nonlinear coupling is above a threshold value~\cite{PRL_self_trap_op,PRL_trapping_op,selftrapping_chen,Cid_2016,Porras_OE_kerr,molina_selftrapp,
prb_moura_nonlinear,prbnonlinear,selftrapp_ACDC}. 

Both scenarious have been widely studied in different areas. In the context of optical fibers, for example, the employment of soliton-like features for optical communications has been studied~\cite{Rev_mod_soliton_op_commu,PRA_op_comm,PRA_op_comm_2,scien_op_comm}. Soliton and soliton-like structures have also been reported as underlying mechanisms of charge carrier transport of conducting polymers~\cite{ssh_prl,rev_mod_phys_ssh,Hegger_rev,ssh_nature,SSH_polaron}. Self-trapped vortex beams azimuthally stable at moderate values of the input intensity have been reported, in which the saturation of the refractive nonlinearity and the instability-suppressing effect of the three-photon absorption display a fundamental role~\cite{Cid_2016}. Driven-dissipative Bose-Einstein condensates (BEC) in a two-mode Josephson system have been used to obtain the alternating-current Josephson effect with magnons, as well as macroscopic quantum self-trapping in a magnon-BEC~\cite{prbnonlinear}. 

Although nonlinear aspects have been reported in the context of discrete-time quantum walks (DTQWs), a full understanding of the phenomenology is still distant. One of the earliest studies that reported  a nonlinear self-phase modulation on the wave function during the walker evolution showed the formation of nondispersive pulses~\cite{GB_DQW}. An anomalous slow diffusion has been reported for a nonlinear quantum walk in which the coin operator depends on the coin states of the nearest-neighbor sites~\cite{SR_nlqw}. The dynamics of a nonlinear Dirac particle has been simulated by using a nonlinear quantum walk, with a description of solitonic behavior and the collisional phenomena between them~\cite{pra_dirac_nl}. By using DTQWs which combine zero modes with a particle-conserving nonlinear relaxation mechanism, a conversion of  two zero modes of opposite chirality into an attractor-repeller pair of the nonlinear dynamics was reported~\cite{topo_nlqw}. 
By investigating the effect of nonlinear spatial disorder on the edge states at the interface between two topologically different regions, the preservation of the ballistic propagation of the walker has been described even for very strong nonlinear couplings~\cite{edgestate_nonde}. Nonlinear effects on the quantum walks ruled by Pauli-X gates homogeneously distributed have been revealed the existence a set of stationary and moving breathers with almost compact superexponential spatial tails~\cite{chaos_nlqw}. Disordered nonlinear DTQWs were used to confirm that the subdiffusive spreading of wave-packets (well-known in Gross-Pitaevskii lattices) persists over an additional four decades, which suggests this subdiffusive behavior as universal~\cite{prl_nldisorder}. Cross-Kerr nonlinearity and orbital angular momentum have been used as two distinct degrees of freedom in the position space, in order to propose a scheme able to perform infinite steps of 2D DTQWs~\cite{osa_conti_kerr}. 

Quantum walks exhibit an exponential superiority over its classical counterpart due to coherent superposition and quantum interference~\cite{Aharonov_qw,quant_info_book}. This feature makes them a versatile tool for the realization of quantum algorithms and quantum simulation~\cite{quant_info_book,qw_search_algo,qw_child}. In this context, studies on quantum computation based on optics are growing, in which the left and right polarization states of a single photon make up a natural computational basis of qubits. Thus, motivated by the wide nonlinear phenomenology in optical systems, we investigate the dynamics of quantum walkers in nonlinear DTQWs. By considering a Kerr-like nonlinearity, we associate an acquisition of the intensity-dependent phase to the walker while it propagates on the lattice. Transport properties are studied by exploring typical quantities such as the inverse participation ratio, the survival probability and the wave-function profile. Adjusting the nonlinear parameter and the quantum gates, we will show the existence of different quantum walking regimes, including those with travelling soliton-like structures or localized by self-trapping. In this latter, the dispersive mode is fully suppressed by nonlinearity, making the walker to be strongly trapped in the initial position developing a breathing mode. This scenario is absent for quantum gates close enough to Pauli-X. It appears for intermediate configurations and becomes predominant as quantum gates get closer to Pauli-Z. The threshold between quantum walks with delocalized or localized regimes exhibits an unusual aspect, in which an increment on the nonlinear strength can induce the system from localized to a delocalized regime. 

\section{model}
\label{model}

In this work, we deal with a quantum walker moving in an infinite 1D nonlinear lattice of interconnected sites. The walker consists of a qubit, whose internal degree of freedom (spin or polarization) determines the direction of movement in discrete steps. Thus, the quantum walker state $|\psi\rangle$ belongs to a Hilbert space $H=H_{c}\otimes H_{p}$, where $H_{c}$ is a complex vector space of dimension 2 associated with the internal degree of freedom of the qubit, and $H_{p}$ denotes a countable infinite-dimensional space associated to lattice sites. We describe the internal degree of freedom spanned by orthonormal basis $\{|R\rangle=(1,0)^{T}$, $|L\rangle=(0,1)^{T}\}$, where the supercript denote transpose, in Hilbert space $H_{c}$. The position space $H_{p}$ is spanned by the orthonormal basis $\{|n\rangle$: $n \in \mathbb{Z}\}$ with $n$ ranging from $n=1$ to $N$. Thus, a general state in the $t$-th time step can be given as 
\begin{equation}
|\psi(t)\rangle=\sum_{n}\left[a(n,t)|R\rangle+b(n,t)|L\rangle\right]\otimes|n\rangle,
\end{equation}
so that the normalization condition is satisfied $\sum_{n}(|a(n,t)|^{2}+|b(n,t)|^{2})=1$. 

In general lines, the dynamical evolution of a discrete-time quantum walk is governed by unitary transformation $|\psi(t)\rangle=\hat{U^{t}}|\psi(t-1)\rangle$, where $\hat{U^{t}}=\hat{S}(\hat{C}\otimes I_{P})$ and $I_{P}$ is the identity operator in space of positions. The conditional shift operator has the form
\begin{eqnarray}
\hat{S}=S_{+}\otimes|R\rangle\langle R|+ S_{-}\otimes|L\rangle\langle L|,
\end{eqnarray}
where $S_{\pm}=\sum_{n=1}^{N}|n\pm 1\rangle\langle n|$, while $\hat{C}$ (well known as quantum coin) is an arbitrary $SU(2)$ unitary operator given by
\begin{eqnarray}
\hat{C}&=& \sum_{n}
[c_{R,R}|R\rangle 
+ c_{R,L}|L\rangle]\langle R| \\ \nonumber
& &+ [c_{L,R}|R\rangle 
- c_{L,L}|L\rangle]\langle L|
\otimes|n\rangle\langle n|,
\end{eqnarray}
with $c_{R,R}=c_{L,L}=\cos(\theta)$, $c_{R,L}=c_{L,R}=\sin(\theta)$. The parameter $\theta\in [0,2\pi]$ controls the variance of the probability distribution of the walk.

Here, in order to introduce the nonlinearity, we add to the dynamical evolution protocol one more operator who describes the acquisition of an intensity-dependent (nonlinear) phase to each of the spinor components~\cite{GB_DQW}. Thus, $\hat{U}^{t}=\hat{S}(\hat{C}\otimes I_{P})\hat{U}_{NL}^{t-1}$ where $\hat{U}_{NL}$ is given by
\begin{eqnarray}
\hat{U}_{NL}^{t}&=&\sum_{s=R,L}\sum_{n=0}^{N} e^{iG^t(n,s)}|s\rangle\langle s|\otimes |n\rangle\langle n|\\ \nonumber
&=&\sum_{n}(e^{iG^t(n,R)}|R\rangle\langle R| + e^{iG^t(n,L)}|L\rangle\langle L|)\otimes|n\rangle\langle n|.
\label{eq4}
\end{eqnarray}
$G^t(n,s)$ is arbitrary function of the probabilities, depending on the internal degree of freedom (coin state) and the lattice site (spacial state). 

Within an optical setup, nonlinearity may be result of nonlinear optical media into the optical paths or the use of detectors on each optical path to measure and control the coin state probability distribution. We consider a Kerr-like nonlinearity, so that $G^t(n,s)=2\pi\chi|\psi_{n,s}^{t}|^{2}$. Here, $\chi$ describes the nonlinear strength of the medium, where the linear discrete-time quantum walk can be recovered by setting $\chi=0$. By using the time-evolution protocol $|\psi(t)\rangle=\hat{U}^{t}|\psi(t-1)\rangle$ we can derive the recursive evolution equations for the probability amplitudes
{\footnotesize
\begin{eqnarray}
\footnotesize
\psi_{n,R}^{t+1}=c_{R,R}e^{i2\pi\chi|\psi_{n+1,R}^{t}|^{2}}\psi_{n+1,R}^{t}+c_{R,L}e^{i2\pi\chi|\psi_{n+1,L}^{t}|^{2}}\psi_{n+1,L}^{t}, \nonumber\\
  \psi_{n,L}^{t+1}=c_{L,R}e^{i2\pi\chi|\psi_{n-1,R}^{t}|^{2}}\psi_{n-1,R}^{t}-c_{L,L}e^{i2\pi\chi|\psi_{n-1,L}^{t}|^{2}}\psi_{n-1,L}^{t}.\nonumber\\
\end{eqnarray}}
Thus, the state of the quantum particle in the $t$-th time step is given by two-component wave-function ($\{\psi_{n,R}^{t},\psi_{n,L}^{t}\}$), where $\psi_{n,R}^{t}$ and $\psi_{n,L}^{t}$ are the probability amplitudes of obtaining the states $|R\rangle$ and $|L\rangle$ at position $n$ and time step $t$, respectively. We consider for the whole analysis open chains as boundary condition,  in which the initial position ($ n_0 $) of the quantum walker is alocated in the central site of the lattice. We stand out that the lattice sizes are large enough so that the wave function does not reached the its edges over the time course described.

\section{Results and discussion}\label{Results_and_discussion}  

\begin{figure}[!t]
\centering 
\includegraphics[width=8.5cm]{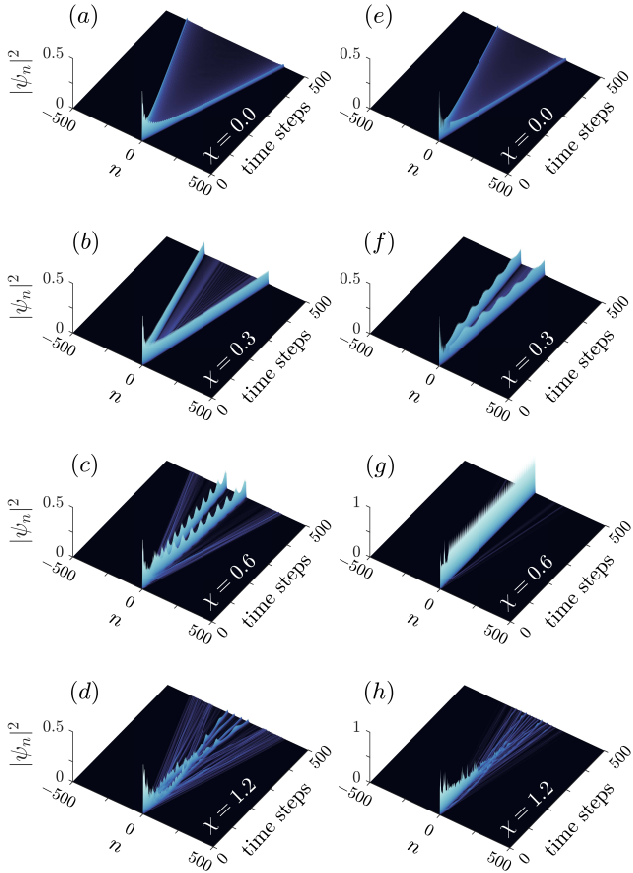}
        \caption{(Color on-line) Time evolution of the density of probability in position space of a quantum walker on chains composed of quantum gates $\theta=\pi/4$ (left column) and $\theta=\pi/3$ (right column) homegeneously distributed. Both quantum gates exhibit travelling soliton-like structures in presence of nonlinearity, whose velocity decreases as nonlinear parameter ($\chi$) increases. Although both scenarious culminates in a scenario of collisions with inelastic scattering for sufficiently strong nonlinearities, a self-trapped quantum walk emerges only for $\theta=\pi/3$, which suggests a phenomenology with gate-dependence.}
        \label{fig1}
\end{figure}

We start following the time evolution of the probability density distribution $|\psi_{n}(t)|^{2}$ for some representative values of the nonlinear parameter $\chi$. With an initial state of walker adjusted as a superposition of left- and right-handed circular polarization ($|\Psi(0)\rangle=1/\sqrt{2}(|R\rangle+i|L\rangle)\otimes|n_0\rangle$), we show in Fig.~\ref{fig1} the dynamics described in chains ruled by quantum gates $\theta=\pi/4$ (left column) and $\theta=\pi/3$ (right column) homegeneously distributed. As expected, in absence of nonlinearity ($\chi=0.0$) both quantum gates induce a spread of the probability distribution through the lattice exhibiting two peaks at the borders of the distribution, whose maximum value monotonically decreases with time. However, this scenario is heavily altered as $\chi$ grows. For $\chi=0.3$, we observe the probability distribution predominatly concentrated in a few lattice positions, by estabilishing two mobile peaks whose size and shape remain approximately constant at time, except for small oscillations around a mean value. Travelling self-trapped states are consistent with the observation of soliton-like structures described in Ref.~\cite{GB_DQW} and have also reported in another systems~\cite{Buarque_nl,Datta1998a}. Results of Hadamard quantum gates ($\theta=\pi/4$) suggest the velocity of the solitons-like formations decreasing as $\chi$ increases, in such way that, for sufficiently strong nonlinearities,  a scenario of collisions with inelastic scattering comes up. However, as we increase the nonlinear parameter for the system ruled by $\theta=\pi/3$, a behavior not previously reported in the literature is observed.  For $\chi=0.6$ the probability distribution remains predominantly trapped around the initial position, i.e. a stationary self-trapped quantum walk. Furthermore, contrary to expectation, the concentration of the walker around the initial position does not grow as $ \chi $ increases. Just like $\theta=\pi/4$, collisions with inelastic scatterings arise for sufficiently strong nonlinearities.

\begin{figure}[!b]
\centering \includegraphics[width=6.3cm]{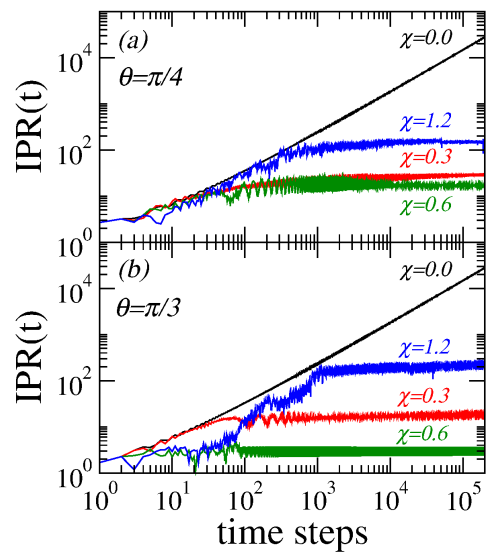}
\caption{(Color on-line) Time evolution of the inverse participation ratio (IPR) for  same configurations used in Fig.~\ref{fig1}. We observe the IPR($t$) recovering relevant aspects reported before, from standard quantum walk ($\chi=0$) to self-trapped quantum walk ($\theta=\pi/3$ with $\chi=0.6$).}
\label{fig2}
\end{figure}

In order to better characterize the previous results we compute de inverse participation ratio,
\begin{eqnarray}
\text{IPR($t$)}=\frac{1}{\sum_{n}|\psi_{n}(t)|^{4}},
\label{IPR}
\end{eqnarray}
that gives the estimate number of lattice sites over which the wavepacket is spread at time $t$. Thus, in Fig.~\ref{fig2} we use the same configurations shown in Fig.~\ref{fig1}, with Fig.~\ref{fig2}a and Fig.~\ref{fig2}b describing the systems ruled by quantum coins $\theta=\pi/4$ and $\theta=\pi/3$, respectively. We observe the IPR($t$) recovering relevant aspects reported before. While the spread of quantum walker is described by a IPR($t$) growing over time in absence of nonlinearity, the dynamics involving soliton-like structures (for $\chi=0.3, 0.6, 1.2$) are described by IPR($t$) approximately constant after an initial transient. The lower inverse participation ratio for $\theta=\pi/3$ and $\chi=0.6$ corroborates the localization induced by self-trapping phenomena described above. On the other hand, multiple collisions between soliton-like structures induce a walker scattering, which explains the behavior of $\chi=1.2$.

We achieve a complementary  analysis by computing the survival probability
\begin{eqnarray}
\text{SP($t$)}=\sum_{s=R,L}|\langle n|\otimes \langle s|\psi(t)\rangle|^2 \bigg|_{n=n_0}. 
\end{eqnarray}
This quantity describes the probability of the walker to be found at the starting position at time $t$. In the long-time regime the survival probability saturates at a finite value for a localized quantum walk, while SP($t\mbox{)} \rightarrow 0$ means the walker escaping from its initial location.

\begin{figure}[!t]
\centering 
\resizebox{8.2cm}{13.5cm}{\includegraphics{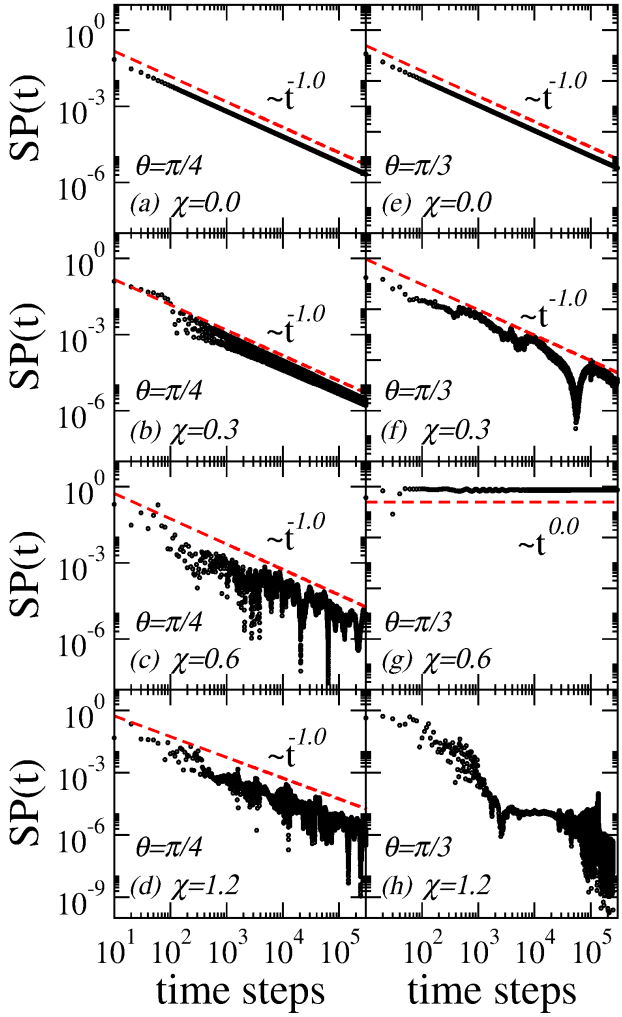}}
        \caption{(Color on-line) Time evolution of the survival probability (SP) for  same configurations used in Fig.~\ref{fig1}. We observe the SP($t$) ratifying all aspects reported before. The scaling behaviour SP($t$)$ \sim t^{-1}$ well defined for almost all configurations, gives way to SP($t$)$\sim t^{0}$ for $\theta=\pi/3$ with $\chi=0.6$, which corroborates a self-trapped quantum walk.}
        \label{fig3}
\end{figure}

In Fig. \ref{fig3} we show the time evolution of SP($t$) for same configurations used before, with left (right) column giving $\theta=\pi/4$ ($\theta=\pi/3$). In absence of nonlinearity, the spreading of walker on the lattice is described by a scaling behaviour SP($t$)$ \sim t^{-1}$, that is in full agreement with explicit expression in ref.~\cite{return_prob}. We also observed this scaling behavior for $\theta=\pi/4$ and $\chi= 0.3, 0.6, 1.2$, which is consistent with soliton-like modes travelling through the lattice, i.e. an absence of the walker localization. On the other hand, for $\theta=\pi/3$ another pattern comes up: for $\chi=0.6$ we have SP($t$)$\sim t^{0}$ after an initial transient, which descibes the walker remaining localized around its initial position. In agreement with Fig.~\ref{fig2}, SP($t$) close to unity for $\chi=0.6$ (see Fig.~\ref{fig3}g) reinforces the ideia of stationary trapping. Besides, SP($t$) decreasing for $\chi=1.2$ confirms an absence of walker localization at the initial site after a long time evolution. The roughness in SP($t$) data suggests destructive interferences of soliton-like structures as time evolves.

\begin{figure}[!b]
\centering 
   \includegraphics[width=8.5cm]{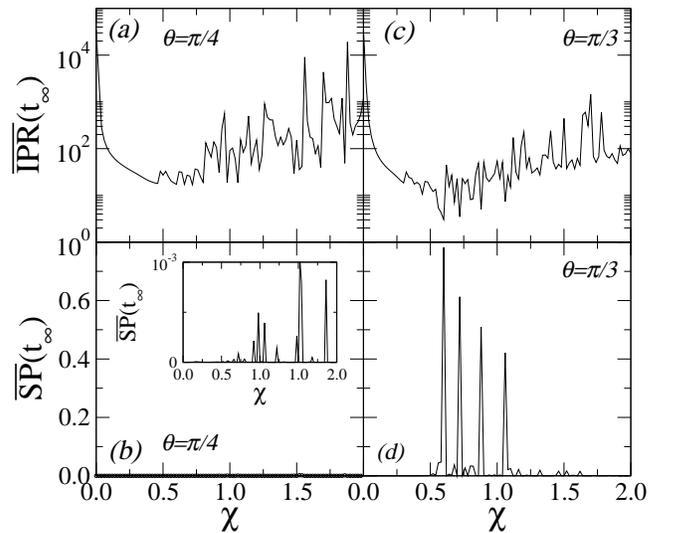}
        \caption{Long-time average of the inverse participation ratio and survival probability versus nonlinear parameter ($\chi$) for quantum gates $\theta=\pi/4$ and $\theta=\pi/3$. Both quantities agree with the existence of travelling soliton-like structures for sufficiently small nonlinearities and with a chaotic-like regime for sufficiently strong nonlinearities. However, the emergence of $\chi$ settings for $\theta=\pi/3$ in which $\overline{\text{IPR}}$($t_\infty\mbox{)}$ and $\overline{\text{SP}}$($t_\infty\mbox{)}$ are close to unity corroborates the quantum walker localization by self-trapping, as well as its gate-dependence.}
        \label{fig4}
\end{figure}

Previous results suggest the regime of localized self-trapped quantum walks as gate-dependent, i.e. restricted to some configurations of quantum gates. This behavior is consistent with the dispersive character associated with the distribution of quantum gates on the lattice~\cite{pre_spectrum}, since the emergence and dynamics of soliton-like structures are associated to balancing between nonlinearity and linear correlation (dispersion,diffraction, diffusion) effects~\cite{nonlinear_optics,physreports_soliton_optics,Davydov_book,Phys_repo_soliton_davydov, rev_mod_phys_soliton,rev_mod_phys_BEC,Holstein_1,Holstein_2,ssh_prl,rev_mod_phys_ssh,Rev_mod_soliton_op_commu,
PRA_op_comm,PRA_op_comm_2,scien_op_comm,Buarque_nl,Hegger_rev,ssh_nature,SSH_polaron}. In order to provide a broader and accurate description, we explore the asymptotic regime of IPR($t$) and SP($t$) for distinct $\theta$ settings. We keep considering infinite lattices, but now we compute an average of both quantities around $10^4$ time steps, identified by $\overline{\text{IPR}}$($t_\infty$) and $\overline{\text{SP}}$($t_\infty$). In Fig.~\ref{fig4} we explore $\theta=\pi/4$ and $\theta=\pi/3$ by ranging the nonlinear parameter $\chi$  between 0 and 2. For the early stage of nonlinearity, both $\overline{\text{IPR}}$($t_\infty$) and $\overline{\text{SP}}$($t_\infty$) suggest delocalized quantum walks in which the walker spreads  on the lattice. Mobile soliton-like structures arise as $\chi$ grows, described by the decrease in $\overline{\text{IPR}}$($t_\infty\mbox{)}$ and $\overline{\text{SP}}$($t_\infty\mbox{)}\sim 0$. The emergence of $\chi$ settings for $\theta=\pi/3$ in which $\overline{\text{IPR}}$($t_\infty\mbox{)}$ and $\overline{\text{SP}}$($t_\infty\mbox{)}$ are close to unity corroborates the quantum walker localization by self-trapping, as well as its gate-dependence. For sufficiently strong nonlinearities, a chaotic-like regime has been found not only for $ \theta = \pi / 4 $~\cite{GB_DQW}, but also for $\theta=\pi/3$. Here, the walker dynamics becomes extremely sensitive to small variations of the nonlinear parameter. This regime comprises quantum walks with delocalized soliton-like structures (where modes are continuously moving apart) and soliton-like dynamics with multiple modes and collisions. This chaotic-like behavior has been shown in other nonlinear systems~\cite{Buarque_nl,Datta1998a}.

\begin{figure}[!t]
\centering 
   \includegraphics[width=8.5cm]{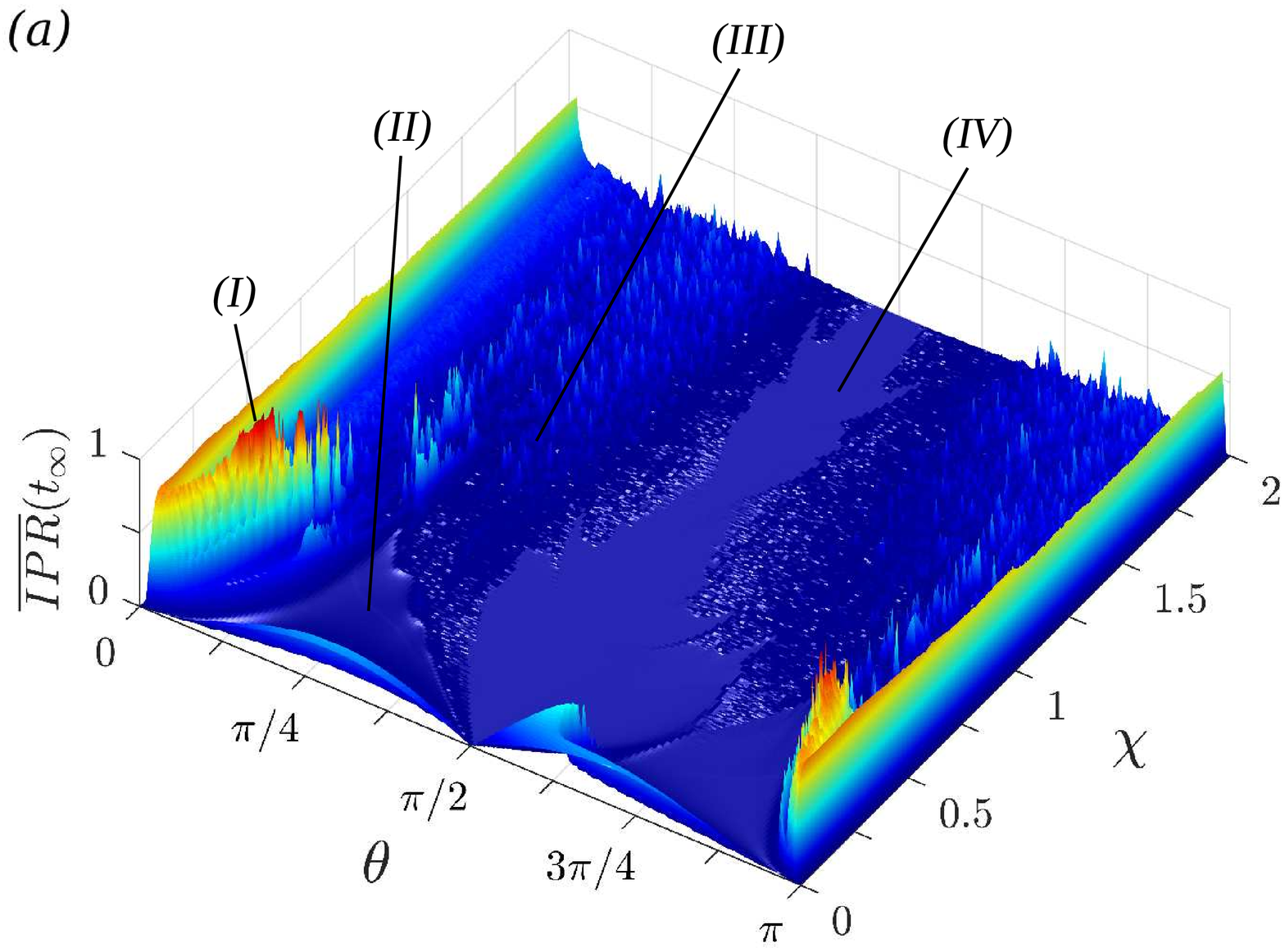}
   \includegraphics[width=8.5cm]{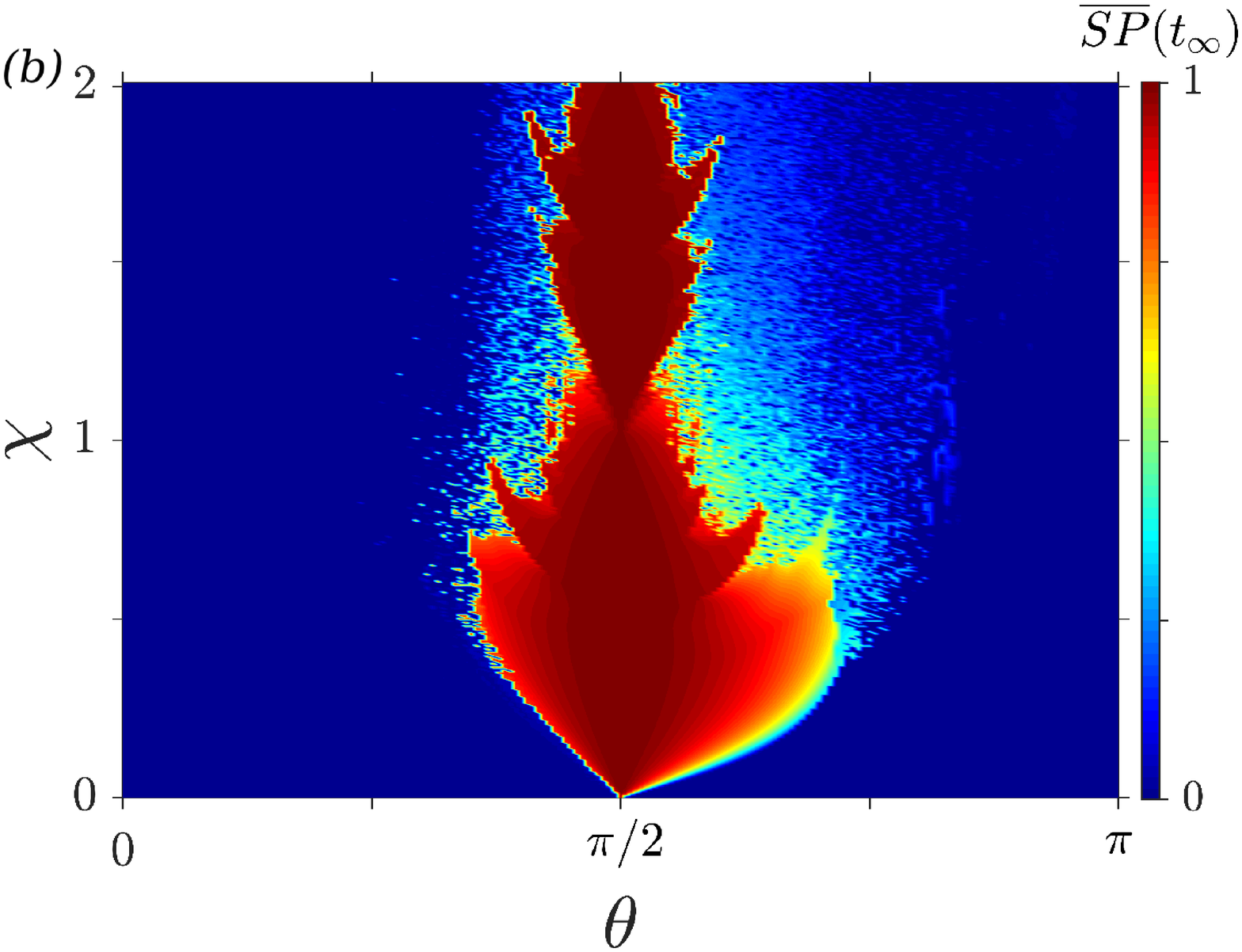}
        \caption{$\chi$ versus $\theta$ diagrams for long-time average of the inverse participation ratio and the survival probability. The initial state of the walker remain described as a superposition of left- and right-handed circular polarization, i.e $ | \Psi (0) \rangle = 1 / \sqrt {2} (| R \rangle + i | L \rangle) \otimes | n_0 \rangle $, with the initial position ($ n_0 $) of the quantum walker allocated in the central site of the lattice. We observe an absence of trapped structures for sufficiently small $\theta$ values, even for strong nonlinear parameter (I). As we increase $\theta$ toward $\pi/2$, different scenarios emerge as we change the $\chi $ value: (II) Soliton-like structures propagating through the lattice;  (III) chaotic-like regime; and (IV) stationary self-trapped quantum walks.}
        \label{fig5}
\end{figure}

For the Fig.~\ref{fig5} we extend our numerical experiments in order to offer $\chi$ versus $\theta$ diagrams. In Fig.~\ref{fig5}a we consider the maximal IPR between collected data in order to plot on the vertical axis a normalized $\overline{\text{IPR}}$($t_\infty$). For Fig.~\ref{fig5}b we compute the $\overline{\text{SP}}$($t_\infty$) as before. The initial state of the walker remain described as a superposition of left- and right-handed circular polarization, i.e $ | \Psi (0) \rangle = 1 / \sqrt {2} (| R \rangle + i | L \rangle) \otimes | n_0 \rangle $, with the initial position ($ n_0 $) of the quantum walker allocated in the central site of the lattice. By simultaneously exploring both diagrams, we observe an absence of trapped structures for sufficiently small $\theta$ values, even for strong nonlinear parameter. In this regime (I), the dispersive character is predominant, with small contribution of the interference terms of $\hat{C}$ matrix. Moreover, in this regime, the increasing on $\overline{\text{IPR}}$($t_\infty$) suggests the nonlinearity as a mechanism able to increase the spread of the walker. However, this behavior is not related to the spreading velocity, but rather to the wave function distribution, that now exhibits a more uniform profile than the one presented in the absence of nonlinearity. The wavefronts exhibit aspects close to a soliton-like strucure, but decrease slowly with time. 

\begin{figure}[!t]
\centering 
   \includegraphics[width=8.5cm]{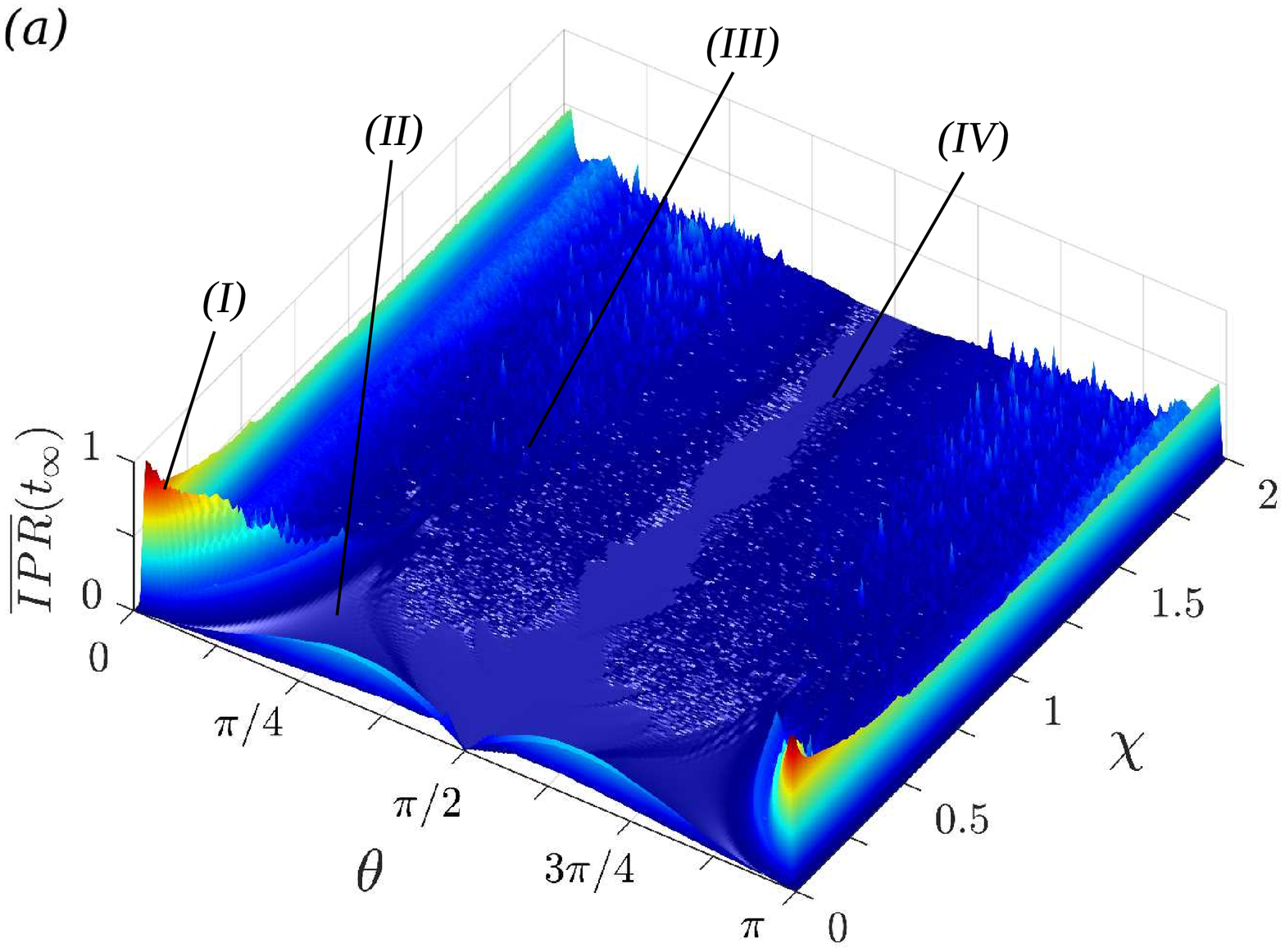}
   \includegraphics[width=8.5cm]{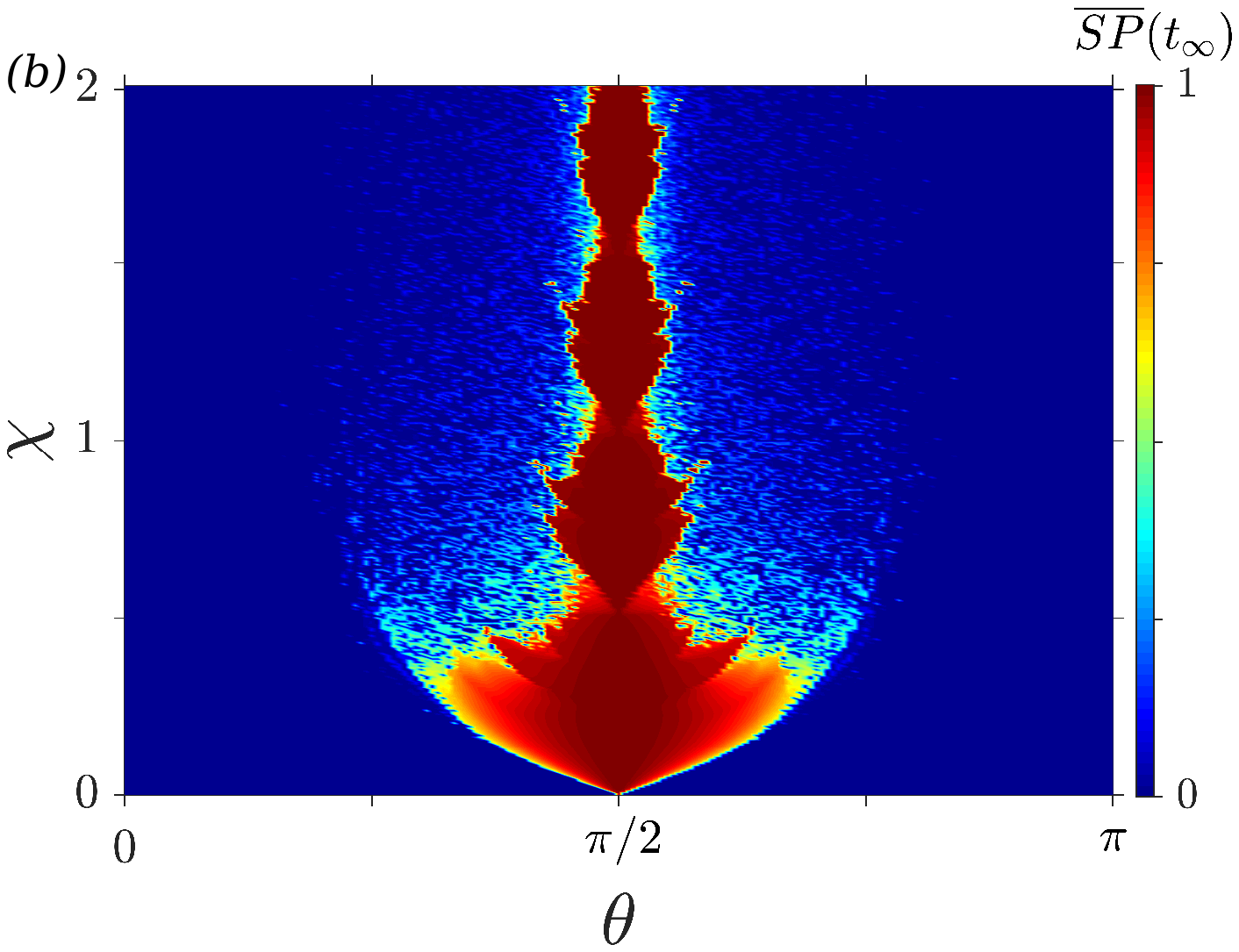}
        \caption{$\chi$ versus $\theta$ diagrams for long-time average of the inverse participation ratio and the survival probability. The initial state of the walker now is given by $ | \Psi (0) \rangle = | R \rangle \otimes | n_0 \rangle $, with the initial position ($ n_0 $) of the quantum walker allocated in the central site of the lattice. In general lines, the phenomenology is homologous to  behavior described by initial state $ | \Psi (0) \rangle = 1 / \sqrt {2} (| R \rangle + i | L \rangle) \otimes | n_0 \rangle $. However, we observe now a symmetric profile around $\theta=\pi/2$ and the regime of stationary self-trapped quantum walks even more concentrated around $\theta=\pi/2$.}
        \label{fig6}
\end{figure}

As we increase $\theta$ toward $\pi/2$, different scenarios emerge as we change the $\chi $ value. For $\chi$ sufficiently small, the normalized $\overline{\text{IPR}}$($t_\infty$)$\sim 0$ with $\overline{\text{SP}}$($t_\infty$)$\sim 0$ is consistent with the existence of soliton-like structures propagating through the lattice (II). As described before (see Fig.~\ref{fig4}), the increasing of $\chi$ promotes a regime in which evolution of the solitons becomes extremely sensitive to small variations of the nonlinear parameter (III). This behavior is found for high enough $\chi$ values and described by fluctations on the normalized $\overline{\text{IPR}}$($t_\infty$) and $\overline{\text{SP}}$($t_\infty$). Stationary self-trapped quantum walks (IV) become evident as we observe the normalized $\overline{\text{IPR}}$($t_\infty$)$\sim 0$ and $\overline{\text{SP}}$($t_\infty$)$\sim 1$. Both diagrams confirm that, once within a stationary self-trapped regime, an increment on $ \chi $ does not mean an increasing of the localization degree. Thus, the threshold between delocalized and localized regimes exhibits an unusual aspect. We also observe the stationary self-trapped regime becoming predominant as $\theta$ get closer to Pauli-Z quantum gates ($\theta=\pi/2$). In this configuration, whose energy spectrum of the two main bands resembles of flat degenerate bands~\cite{pre_spectrum}, the dynamics who takes $|R\rangle$ to $|L\rangle$ and $|L\rangle$ to $|R\rangle$  is reinforced by nonlinear (probability-dependent) phase.

A character absent in previous discussion is the asymmetric aspect of the normalized $\overline{\text{IPR}}$($t_\infty$) and $\overline{\text{SP}}$($t_\infty$) diagrams around Pauli-Z quantum gates. This behavior is associated to complex component on the left-handed circular polarization of initial state of walker, which gives opposite signals for $\theta <\pi/2$ and $\theta>\pi/2$ for the dynamical evolution protocol due to acquisition of an intensity-dependent (nonlinear) phase described in eq.~\ref{eq4}. This statement becomes more evident when we show results obtained by employing the same methodology used earlier on the condition in which the initial state of the walker is given by $ | \Psi (0) \rangle = | R \rangle \otimes | n_0 \rangle $ (see Fig.~\ref{fig6}). In general lines, the phenomenology is homologous, with normalized $\overline{\text{IPR}}$($t_\infty$) and $\overline{\text{SP}}$($_\infty$) diagrams exhibiting the same regimes, but with the symmetric profile around $\theta=\pi/2$. Now, the regime of stationary self-trapped quantum walks is even more concentrated around $\theta=\pi/2$.

\section{Summary and conclusions}

In summary, we have studied the dynamics of quantum walkers in nonlinear DTQWs. By consider a Kerr-like nonlinearity, we associate an acquisition of the intensity-dependent phase to the walker while it propagates on the lattice. With nonlinear strength of the medium as an adjustable parameter, we explored the role of quantum gates on the emergence of mobile soliton-like structures and quantum walker dynamics, as well as the regime in which the quantum walker exhibits a localization induced by self-trapping.  In this latter the dispersive mode is fully suppressed by nonlinearity, making the walker to be strongly trapped in the initial position developing a breathing mode. The stationary self-trapped regime becomes predominant as $\theta$ get closer to Pauli-Z quantum gates, we also have shown that the threshold between delocalized and localized regimes exhibits an unusual aspect, in which an increment on the nonlinear parameter can induce the system of a localized to a delocalized regime. To conclude, by considering that nonlinearity has attracted much attention in quantum information science~\cite{comput_kerr,prl_crosskerr,nonlinear_phasegates,osa_conti_kerr}, we hope that our work may impel further investigations on quantum walks in nonlinear optical media. Under experimental point of view, we consider optical systems as the most promising in the implementation of our studies, in which we suggest the use of nonlinear optical media into the optical paths or detectors to measure and control the coin state probability distribution.

\section{Acknowledgments}

This work was partially supported by CNPq (The Brazilian National Council for Scientific and Technological Development), CAPES (Federal Brazilian Agency) and FAPEAL (Alagoas State Agency).

\bibliographystyle{nature}
\bibliography{referencias}

\begin{thebibliography}{10}

\bibitem{nonlinear_optics}
Hennig, D. and Tsironis, G.
\newblock {\em Physics Reports}{ \bf 207}, 333--432 (1999).

\bibitem{Review_mod_phy_nonlinear_optics}
M.~Fleischhauer, A.~Imamoglu, J. P.~M.
\newblock {\em Review of Modern Physics}{ \bf 77}, 633 (2005).

\bibitem{physreports_soliton_optics}
Lederer, F., Stegeman, G.~I., Christodoulides, D.~N., Assanto, G., Segev, M.,
  and Silberberg, Y.
\newblock {\em Physics Reports}{ \bf 463}, 1--126 (2008).

\bibitem{Davydov_book}
Davydov, A.~S.
\newblock {\em Solitons in Molecular Systems}.
\newblock Mathematics and Its Applications. Springer Netherlands,  (1985).

\bibitem{Phys_repo_soliton_davydov}
Scott, A.
\newblock {\em Physics Reports}{ \bf 217}, 1--67 (1992).

\bibitem{rev_mod_phys_Pitaevskii}
Dalfovo, F., Giorgini, S., Pitaevskii, L.~P., and Stringari, S.
\newblock {\em Reviews of Modern Physics}{ \bf 71}, 463 (1999).

\bibitem{rev_mod_phys_soliton}
Kartashov, Y.~V., Malomed, B.~A., and Torner, L.
\newblock {\em Reviews of Modern Physics}{ \bf 83}, 247 (2011).

\bibitem{rev_mod_phys_BEC}
Morsch, O. and Oberthaler, M.
\newblock {\em Rev. Mod. Phys.}{ \bf 78}, 179 (2006).

\bibitem{Holstein_1}
Holstein, T.~D.
\newblock {\em Annals of Physics}{ \bf 8}, 325 (1959).

\bibitem{Holstein_2}
Holstein, T.~D.
\newblock {\em Annals of Physics}{ \bf 8}, 343 (1959).

\bibitem{ssh_prl}
Su, W.~P., Schrieffer, J.~R., and Heeger, A.~J.
\newblock {\em Physical Review Letters}{ \bf 42}, 1698 (1979).

\bibitem{rev_mod_phys_ssh}
Heeger, A.~J., Kivelson, S., Schrieffer, J.~R., and Su, W.~P.
\newblock {\em Reviews of Modern Physics}{ \bf 60}, 781 (1988).

\bibitem{Rev_mod_soliton_op_commu}
Haus, H.~A. and Wong, W.~S.
\newblock {\em Reviews of Modern Physics}{ \bf 68}, 423--444 (1996).

\bibitem{PRA_op_comm}
Kruglov, V.~I. and Harvey, J.~D.
\newblock {\em Physical Review A}{ \bf 98}, 063811 (2018).

\bibitem{PRA_op_comm_2}
Hause, A., Mahnke, C., and Mitschke, F.
\newblock {\em Physical Review A}{ \bf 98}, 033814 (2018).

\bibitem{scien_op_comm}
Kippenberg, T.~J., Gaeta, A.~L., Lipson, M., and Gorodetsky, M.~L.
\newblock {\em Science}{ \bf 361}, 567 (2018).

\bibitem{Buarque_nl}
Buarque, A. R.~C. and Dias, W.~S.
\newblock {\em Communications in Nonlinear Science and Numerical Simulation}{
  \bf 63}, 365 (2018).

\bibitem{Hegger_rev}
Heeger, A.~J.
\newblock {\em Reviews of Modern Physics}{ \bf 73}, 681 (2001).

\bibitem{ssh_nature}
Meier, E.~J., An, F.~A., and Gadway, B.
\newblock {\em Nature Communications}{ \bf 7}(13986), 5382 (2016).

\bibitem{SSH_polaron}
J.~J.~Liu, a. Z. J.~W., Zhang, Y.~L., Meng, Y., and Di, B.
\newblock {\em Journal of Physical Chemistry B}{ \bf 121}, 2366 (2017).

\bibitem{PRL_self_trap_op}
Chiao, R.~Y., Garmire, E., and Townes, C.~H.
\newblock {\em Phys. Rev. Lett.}{ \bf 13}, 479 (1964).

\bibitem{PRL_trapping_op}
Mingaleev, S.~F. and Kivshar, Y.~S.
\newblock {\em Physical Review Letters}{ \bf 86}, 5474 (2001).

\bibitem{selftrapping_chen}
Chen, Y.
\newblock {\em Optics Letters}{ \bf 16}, 4 (1991).

\bibitem{Cid_2016}
Reyna, A.~S., Boudebs, G., Malomed, B.~A., and de~Ara\'ujo, C.~B.
\newblock {\em Phys. Rev. A}{ \bf 93}, 013840 (2016).

\bibitem{Porras_OE_kerr}
Porras, M.~A.
\newblock {\em Optics Express}{ \bf 26}, 19606 (2018).

\bibitem{molina_selftrapp}
Bustamante, C.~A. and Molina, M.~I.
\newblock {\em Phys. Rev. B}{ \bf 62}, 15287 (2000).

\bibitem{prb_moura_nonlinear}
Dias, W.~S., Lyra, M.~L., and de~Moura, F. A. B.~F.
\newblock {\em Phys. Rev. B}{ \bf 82}, 233102 (2010).

\bibitem{prbnonlinear}
Nakata, K., van Hoogdalem, K.~A., Simon, P., and Loss, D.
\newblock {\em Physical Review B}{ \bf 90}, 144419 (2014).

\bibitem{selftrapp_ACDC}
Pereira, A. L.~S., Lyra, M.~L., de~Moura, F. A. B.~F., Neto, A.~R., and Dias,
  W.~S.
\newblock {\em Communications in Nonlinear Science and Numerical Simulation}{
  \bf 64}, 89 (2018).

\bibitem{GB_DQW}
Navarrete-Benlloch, C., P\'erez, A., and Rold\'an, E.
\newblock {\em Physical Review A}{ \bf 75}, 062333 (2007).

\bibitem{SR_nlqw}
Shikano, Y., Wada, T., and Horikawa, J.
\newblock {\em Scientific Reports}{ \bf 4}, 4427 (2014).

\bibitem{pra_dirac_nl}
Lee, C.-W., Kurzy\ifmmode~\acute{n}\else \'{n}\fi{}ski, P., and Nha, H.
\newblock {\em Physical Review A}{ \bf 92}, 052336 (2015).

\bibitem{topo_nlqw}
Gerasimenko, Y., Tarasinski, B., and Beenakker, C. W.~J.
\newblock {\em Physical Review A}{ \bf 93}, 022329 (2016).

\bibitem{edgestate_nonde}
Verga, A.~D.
\newblock {\em The European Physical Journal B}{ \bf 90}, 41 (2017).

\bibitem{chaos_nlqw}
Vakulchyk, I., Fistul, M.~V., Zolotaryuk, Y., and Flach, S.
\newblock {\em Chaos}{ \bf 28}, 123104 (2018).

\bibitem{prl_nldisorder}
Vakulchyk, I., Fistul, M.~V., and Flach, S.
\newblock {\em Physical Review Letters}{ \bf 122}, 040501 (2019).

\bibitem{osa_conti_kerr}
Gao, W.-C., Cao, C., Liu, X.-F., Wang, T.-J., and Wang, C.
\newblock {\em OSA Continuum}{ \bf 2}, 1667 (2019).

\bibitem{Aharonov_qw}
Aharonov, Y., Davidovich, L., and Zagury, N.
\newblock {\em Phys. Rev. A}{ \bf 48}, 1687 (1993).

\bibitem{quant_info_book}
Nielsen, M.~A. and Chuan, I.~L.
\newblock {\em Quantum Computation and Quantum Information}.
\newblock Cambridge University Press,  (2000).

\bibitem{qw_search_algo}
Shenvi, N., Kempe, J., and Whaley, K.~B.
\newblock {\em Phys. Rev. A}{ \bf 67}, 052307 (2003).

\bibitem{qw_child}
Childs, A.~M.
\newblock {\em Phys. Rev. Lett.}{ \bf 102}, 180501 (2009).

\bibitem{Datta1998a}
Datta, P.~K. and Jayannavar, A.~M.
\newblock {\em Phys. Rev. B}{ \bf 58}, 8170--8173 Oct  (1998).

\bibitem{return_prob}
Xua, X.-P.
\newblock {\em Eur. Phys. J. B}{ \bf 77}, 479--488 (2010).

\bibitem{pre_spectrum}
Buarque, A. R.~C. and Dias, W.~S.
\newblock {\em Physical Review E}{ \bf 100}, 032106 (2019).

\bibitem{comput_kerr}
Shapiro, J.~H.
\newblock {\em Physical Review A}{ \bf 73}, 062305 (2006).

\bibitem{prl_crosskerr}
Brod, D.~J. and Combes, J.
\newblock {\em Physical Review Letters}{ \bf 117}, 080502 (2016).

\bibitem{nonlinear_phasegates}
Nysteen, A., McCutcheon, D. P.~S., Heuck, M., M\o{}rk, J., and Englund, D.~R.
\newblock {\em Physical Review A}{ \bf 95}, 062304 (2017).

\end{thebibliography}

\end{document}